# Dirac Electrons in a Dodecagonal Graphene Quasicrystal

Sung Joon Ahn[1†], Pilkyung Moon[2†], Tae-Hoon Kim[3†], Hyun-Woo Kim[1], Ha-Chul Shin[1], Eun Hye Kim[1], Hyun Woo Cha[3], Se-Jong Kahng[4], Philip Kim[5], Mikito Koshino[6], Young-Woo Son[7]*, Cheol-Woong Yang[3]* and Joung Real Ahn[1,8]*

[1]*Department of Physics and SAINT, Sungkyunkwan University, Suwon, Republic of Korea*

[2]*New York University and NYU-ECNU Institute of Physics at NYU Shanghai, Shanghai, China*

[3]*School of Advanced Materials Science and Engineering, Sungkyunkwan University, Suwon, Republic of Korea*

[4]*Department of Physics, Korea University, Seoul, Republic of Korea*

[5]*Department of Physics, Harvard University, Cambridge, MA, USA*

[6]*Department of Physics, Osaka University, Machikaneyama, Toyonaka, Japan*

[7]*Korea Institute for Advanced Study, Seoul, Republic of Korea*

[8]*Samsung-SKKU Graphene Center, Sungkyunkwan University, Suwon, Republic of Korea*

[*]To whom correspondence should be addressed.

E-mail: jrahn@skku.edu(JRA), cwyang@skku.edu(CWY), hand@kias.re.kr (YWS)

[†]These authors contributed equally to this work.

**ABSTRACT**

Quantum states of quasiparticles in solids are dictated by symmetry. Thus, a discovery of unconventional symmetry can provide a new opportunity to reach a novel quantum state. Recently, Dirac and Weyl electrons have been observed in crystals with discrete translational symmetry. Here we experimentally demonstrate Dirac electrons in a two-dimensional quasicrystal without translational symmetry. A dodecagonal quasicrystal was realized by epitaxial growth of twisted bilayer graphene rotated exactly 30°. The graphene quasicrystal was grown up to a millimeter scale on SiC(0001) surface while maintaining the single rotation angle over an entire sample and was successfully isolated from a substrate, demonstrating its structural and chemical stability under ambient conditions. Multiple Dirac cone replicated with the 12-fold rotational symmetry were observed in angle resolved photoemission spectra, showing its unique electronic structures with anomalous strong interlayer coupling with quasi-periodicity. Our study provides a new way to explore physical properties of relativistic fermions with controllable quasicrystalline orders.

**ONE SENTENCE SUMMARY:** A Dirac fermion quasicrystal with 12-fold rotational symmetry and without any translational symmetry can be realized from twisted bilayer graphene rotated exactly 30°.

Electronic structures of solids can be controlled by symmetry and spatial dimension of their lattices as well as topological numbers. Discoveries of new symmetries or topological orders in solids have revealed novel quantum states associated with the orders. There are certain structures of quasiperiodic orders without a spatial periodicity essential to typical crystals (*1-3*). Such a quasicrystal was first observed using an Al-Mn alloy in 1984 by D. Shechtman et al (*1*), and have appeared in various intermetallic alloys and several artificial systems (*3-11*). The quasicrystals have been used to study quantum states in quasiperiodic order in comparison to the states in periodic order and disorder, where the quasiperiodic order is considered as the intermediate between the two limits (*12, 13*). To understand the influence of quasiperiodic order on extended wave functions of ordered states, pseudogaps, fine structures of density of states and resonant states have been mainly focused on (*14-22*). However, so far the studies on the quantum states of quasicrystals have been limited to non-relativistic fermions.

Here we report experimentally that the relativistic Dirac fermion quasicrystal can be realized when the Dirac electrons in a single-layer graphene are incommensurately modulated by another single-layer graphene which is rotated by an exact 30°. Such a twisted bilayer graphene (TBG) can form a 12-fold quasicrystal through interlayer interactions between the two layers. This TBG quasicrystal is a direct material realization of the algorithmic quasicrystal construction process first proposed by Stampfli using two ideal hexagonal grids rotated 30° with respect to each other (*23-26*). We observe that the graphene quasicrystal has infinitely many distinct Dirac cone replicas in the Brillouin zone. Furthermore, the graphene quasicrystal shows characteristic electronic structures that differ from the periodic TBG crystals with rotation angles close to 30°. Angle-resolved photoemission spectroscopy (ARPES) clearly exhibits many replicas of Dirac cones with 12-fold rotational symmetry, indicating large enhancement of interlayer coupling with the quasicrystalline order. We show that the graphene quasicrystal can be an excellent platform to explore quantum states of Dirac electrons in quasiperiodic order, just like a single layer graphene has served as the quintessential materials for studying Dirac fermions in periodic crystals. Furthermore, it has been also predicted theoretically that four dimensional quantum hall effects of Dirac fermions can be studied in the two-dimensional quasicrystal (*27*). Therefore, the experimental achievement of the quasicrystal of Dirac fermions in two-dimension can be a starting point to approach higher dimensional physics of quasiparticles in solids.

The graphene quasicrystal was grown on a Si-face of 4H-SiC(0001) surface (see Fig. S1 in Supplementary Information (SI)). The upper and lower layer graphene of the TBG have rotational angles of 0° and 30° with respective to the orientation of the SiC (0001) surface (hereafter briefly referred as R0° and R30°), respectively. Usually, when graphene is thermally grown on a Si-face of 4H- or 6H-SiC(0001), the graphene takes R30° configuration (*28-31*). In contrast, by using borazine gas in a vacuum system, a hexagonal boron nitride (h-BN) layer with R0° can grow epitaxially at 1050 °C. So, we first grow h-BN layer with R0° and then heated the system at a higher temperature of 1600 °C to replace h-BN by graphene layer with maintaining the same angle of 0°. This procedure enable formation of the upper layer of the graphene quasicrystal (*32*). Finally, when the sample was heated further at 1600 °C, a graphene layer with R30° grows between the first graphene layer with R0° and the SiC (0001) substrate, resulting in twisted bilayer graphene with an exact rotational angle of 30°. The epitaxial growth has a merit over a manual transfer (*33-36*) in making the quasicrystal because of the inevitable errors in twisting angles as well as the local distortions during the transfer process.

We determined the crystal orientation of the graphene layers from low energy electron diffraction (LEED) and transmission electron microscopy (TEM) measurements (Fig. 1). Figure 1A shows a LEED pattern of the graphene quasicrystal. The upper and lower layers of the TBG can be clearly distinguished from the LEED pattern, since the intensity of the 1×1 LEED pattern of the upper layer graphene (the red hexagon in Fig. 1A) is higher than that of the lower layer graphene (the blue hexagon in Fig. 1A). The LEED spots other than the 1×1 LEED patterns come from the quasiperiodic order induced by the interlayer interaction. As shown in Fig. 1B, our observation is fully consistent with the simulated pattern from an atomic structure model of graphene quasicrystal shown in Fig. S2 in SI. By using the 1×1 mm$^2$ sized electron beam, we also confirmed that the LEED pattern is uniform throughout the entire sample size of up to 4×7 mm$^2$, where the sample size is limited only by our experimental set-up and can be scaled up to a wafer scale.

The graphene quasicrystal can be spatially mapped onto a quasicrystal lattice model constructed by dodecagonal compound tessellations (see Figs. 1D-E and Figs. S3D-E in SI) (*23-26*). Squares, rhombuses, equilateral triangles with different orientations can fill the entire space with a 12-fold rotationally symmetric pattern without translational symmetry. Since the Stampfli tiles have a fractal structure with self-similarity, the same pattern emerges at a larger scale with an irrational scaling factor. For the graphene quasicrystal, the Stampfli tiles have the scaling factor of $\sqrt{2+\sqrt{3}}$ (Figs. 1D-E and Fig. S3 in SI) (*23*). As shown in the atomic model (Figs. 1C-D), the graphene quasicrystal results in Stampfli tiles such as equilateral triangles and rhombuses. In the false colored TEM image of the graphene quasicrystal transferred from a SiC wafer to a TEM grid (Fig. 1F and Fig. S3D in SI), the Stampfli tiles including squares (blue), rhombuses (red), equilateral triangles (green) with different orientations were obviously observed. The LEED pattern and TEM image clearly support that the twisted bilayer graphene with a rotational angle of 30° has a quasicrystalline order with 12-fold rotational symmetry. We note that, for TEM experiments, the graphene quasicrystal grown on a SiC wafer should be transferred on other substrates and should be robust to chemical treatments in air. The successful TEM experiments support that graphene quasicrystal can be isolated from a substrate and are chemically and structurally stable at room temperature in air. We expect that the robustness of graphene quasicrystal can lead to further studies on the physical properties and applications.

Having established quasicrystalline order in our sample, now we discuss its characteristic electronic structures based on ARPES measurements (Fig. 2). Figure 2A shows the constant energy map of ARPES spectra of graphene quasicrystal, and Fig. 2C shows the enlarged view in the vicinity of the Γ point (see Fig. S4 in SI also). Several replicas of main Dirac cones respecting the 12-fold rotational symmetry are clearly observed throughout the momentum space (Fig. 2) in sharp contrast to the usual TBGs with an arbitrary rotational angle other than 30° (*36-39*). Considering the large photon beam size of 3×3 mm$^2$ in our ARPES, the clear energy-momentum dispersions throughout the entire momentum space ensure the uniform sample quality as in the LEED experiments. Like the LEED patterns, the intensities of the photoemission spectra of the lower graphene layer is weaker than that of the upper graphene layer (Fig. 2A) because of attenuated intensities for photo-emitted electrons from the lower layer.

To understand origins of the 12-fold Dirac cone replicas, we performed a numerical simulation of the ARPES spectra using a single-particle tight-binding model that has been successfully used for describing various electronic properties of TBGs of general angles (*40*) (see the SI for the detailed methods). Figure 3A shows the constant energy map at the Fermi energy (i.e. 0.3 eV above the Dirac point) of simulated ARPES spectra. The red (blue) contours indicate that those Dirac cones mainly originate from inter- and intra-layer scatterings of the wave functions of the upper (lower) graphene layer. In stacked crystals, a state $|\mathbf{k}\rangle$ in the upper layer is scattered to every $|\mathbf{k'}\rangle$ state in the lower layer which satisfies the generalized Umklapp scattering condition of $\mathbf{k}+\mathbf{G}=\mathbf{k'}+\mathbf{G'}$. Here, $\mathbf{k}$ and $\mathbf{G}$ ($\mathbf{k'}$ and $\mathbf{G'}$) are the in-plane wave vector and reciprocal vector of the upper (lower) layer, respectively (*41, 42*). The coupling strength between $|\mathbf{k}\rangle$ and $|\mathbf{k'}\rangle$ is proportional to $t(\mathbf{k}+\mathbf{G}) = t(\mathbf{k'}+\mathbf{G'})$, where $t(\mathbf{q})$ is the in-plane Fourier transform of the interlayer transfer integral(*41*). Figures 3B-D show some examples of the Umklapp scattering from the Dirac point ($\mathbf{k} = \mathbf{K}$) of the upper layer, which clearly shows that the scattering with different $\mathbf{G}$ are mapped to distinct wave vectors $\mathbf{k'}$ in the lower layer (see SI for detailed explanations) (*43*). Each scattering makes the replica of the Dirac cone (originally at $\mathbf{k} = \mathbf{K}$) at $\mathbf{k'}$. Since $\mathbf{G}$ and $\mathbf{G'}$ are neither identical nor commensurate in graphene quasicrystal, there are infinitely many distinct Dirac cone replicas in the momentum space. Likewise, the Dirac cone at $\mathbf{k'}$ of the lower layer is replicated at $\mathbf{k}$ of the upper layer satisfying $\mathbf{k}+\mathbf{G}=\mathbf{k'}+\mathbf{G'}$. From these calculations, the observed positions of Dirac cones in our ARPES experiment are reproduced in our simulations as shown in Figs. 3E-F. The observed positions, however, cannot be reproduced if the angle deviates 30° very slightly (see Figs. S4D-E in SI). For example, if the twisted angle is 29.9576°, the TBG has a translational symmetry with a periodic supercell that is 1,351 times of the graphene unit cell and the computed positions of the Dirac cone replicas in this quasicrystal approximant is not consistent with the experimental ones (Fig. S4E in SI).

Close inspection of ARPES intensities of Dirac cone replicas highlights the important role of higher order Umklapp scatterings in realizing twelve-fold Dirac cones in graphene quasicrystals. Within a single Umklapp scattering process, the scattering strength is proportional to the wave vector $\mathbf{q}$ component of the in-plane Fourier transform of the interlayer transfer integral $t(\mathbf{q})$, where $\mathbf{q}$ ($\equiv \mathbf{k}+\mathbf{G} = \mathbf{k'}+\mathbf{G'}$) is the equivalent wave vector common to both layers. Figures 3B-D show the examples of the scattering with the shortest three $|\mathbf{q}|$'s, and Fig. 3E shows the entire map of the Dirac cone replicas up to the shortest eight $|\mathbf{q}|$. The numbers in the circles indicate the order of the length of $|\mathbf{q}|$ (see SI and Fig. S8)(*41, 42*). If considering a single Umklapp scattering, the scattering intensity should decrease rapidly with $|\mathbf{q}|$ as shown in Fig. 3G because $t(\mathbf{q})$ decays exponentially with large $|\mathbf{q}|$ as demonstrated in Fig. S8C in SI. However, since a number of Umklapp scattering also increases very rapidly with increasing $|\mathbf{q}|$, the intensities of higher order Dirac cone replica increases enormously compared with single scattering intensities as shown in Fig. 3G (e.g., $1.15 \times 10^5$ and $7.40 \times 10^{11}$ times enhancements of intensities with $|\mathbf{q}|$=6.14 Å$^{-1}$ and 9.01 Å$^{-1}$, respectively)**.** Therefore, the resulting intensities for larger $|\mathbf{q}|$ becomes to be more or less similar to each other as shown in Fig. 3G. In experiment, we observe a similar trend in ARPES intensities as shown in Fig. 3H. The first two intensities of Dirac cone replica decrease exponentially and the remaining Dirac cones show almost constant intensities. Although a general trend of intensities in our experiment agrees well with theoretical expectations, the absolute relative intensities do not match the computed ones. The discrepancies between experimental and simulated intensities of the Dirac cone replicas may originates from hidden scattering paths such as the surface reconstruction of SiC substrate (*28, 30, 44-46*) or impurities that may not be properly accounted in the simulation based on the single-particle pictures.

Detailed shapes of Dirac cone replica in the graphene quasicrystal also show their characteristic features. They are all electron-doped and the Dirac point energies of replica on the upper and lower layer are nearly identical (Figs. 4A-B). In typical epitaxial graphene on a SiC (0001) substrate, electrons are transferred from the SiC substrate to graphene (*28-30, 44*). Likewise, we expect the relative shift of the Dirac points between the two layers due to the effective perpendicular electric field as shown in recent studies on gated TBGs (*47, 48*). Unlike gated TBGs, however, all Dirac cones in the sample have almost identical Dirac point energies implying the strong interlayer coupling or larger interlayer screening. This anomalous interlayer screening is in accordance with the exponential enhancement of ARPES intensities of high order Dirac cone replica (Figs. 3G-H) thanks to the higher order Umklapp scatterings. Regardless of the large variation in ARPES intensities of Dirac cone replica, their Fermi velocities are almost constant (Figs. 4C and see S4F in SI also) (*49, 50*). Overall interaction strength between Dirac cones on upper and lower layer can be estimated as $|t(\mathbf{q})/\Delta E|^2$ where $\Delta E$ is the energy difference between $|\mathbf{k}\rangle$ and $|\mathbf{k}^\varnothing\rangle$ states. In TBGs, $\Delta E$ for the dominant interaction near the Dirac point is proportional to the rotation angle. Thus, TBGs with small angles exhibit very strong coupling between $|\mathbf{k}\rangle$ and $|\mathbf{k}^\varnothing\rangle$, since $\Delta E$ is very small, giving a strong distortion to the Dirac cone structures such as the additional renormalization of Fermi velocity (*40, 51*). In graphene quasicrystal, however, $\Delta E$ is very large so that we expect negligible distortion to the Dirac cone structures. As expected, all the Dirac cones show well-defined linear momentum-energy relationship as shown in Figs. 4B-C. Furthermore, the Fermi velocities of the Dirac cones are not quite different from each other as expected from the fact that the Umklapp scatterings involving with large |**q**| keep nearly the same Fermi velocity.

To conclude, a dodecagonal graphene quasicrystal is synthesized for the first time and the robust quasicrystalline order is maintained up to the millimeter scale size. The quasicrystal composed of two graphene layers are shown to be quite stable against several mechanical and chemical treatments. A distinct 12-fold rotational symmetric Dirac electrons are observed and generation of quasicrystalline orders of Dirac electrons is attributed to the incommensurate Umklapp scatterings between the two layers. Further theory is required to explain few spectral features of graphene quasicrystals that are not captured well considering an ideal graphene quasicrystal. Our observations of a dodecagonal quasicrystal using twisted bilayer hexagonal lattices suggest a new way to construct and study quantum phenomena with quasicrystalline order by stacking various two dimensional crystals.

**Acknowledgments:** This work was supported by the National Research Foundation (NRF) of Korea (NRF-2015R1A2A2A01004853 and NRF-2017M2A2A6A01019384). M.K. was supported by JSPS KAKENHI Grant Numbers JP25107005 and JP15K21722. Y.-W.S was supported by NRF of Korea (Grant No.2017R1A5A1014862, SRC program: vdWMRC center). P.M. was supported by the NYU Shanghai (Start-Up Funds), NYU-ECNU Institute of Physics and the NSF of China (Grant No. 11550110177, Research Fund for International Young Scientists program). P.K. was supported by Global Research Laboratory Program (2015K1A1A2033332) through NRF of Korea. C.W.Y. was supported by the National Research Council of Science and Technology (NST) (CRC-15-06-KIGAM) and NRF (NRF-2011-0030058, NRF-2015R1D1A1A01059653) of Korea.

**Supplementary Information:**
Materials and Methods
References (*R1-R9*)
Figures S1-S10
Sections 1-2

# FIGURES

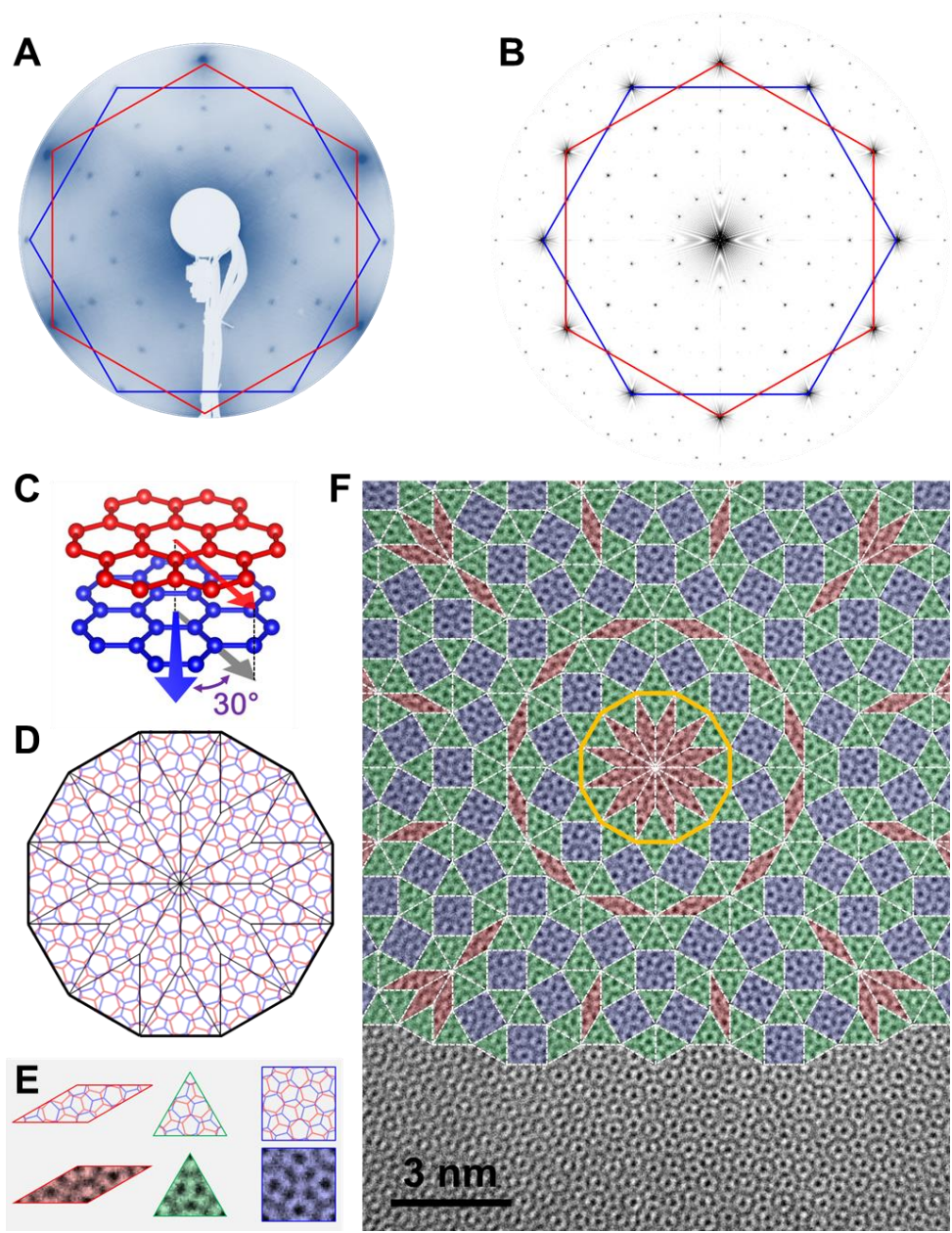

**FIGURE 1**. **A LEED pattern and a TEM image of graphene quasicrystal.** (A) A LEED pattern of graphene quasicrystal. (B) A Fourier transformed pattern of graphene quasicrystal (see also Fig. S2 in SI). (C)-(D) An atomic structure model of twisted bilayer graphene with a rotational angle of 30°. (E) Atomic structures and TEM images of Stampfli tiles (rhombuses (red), equilateral triangles (green), squares (blue)). (F) A false colored TEM image of graphene quasicrystal mapped with 12-fold Stampfli-inflation tiling.

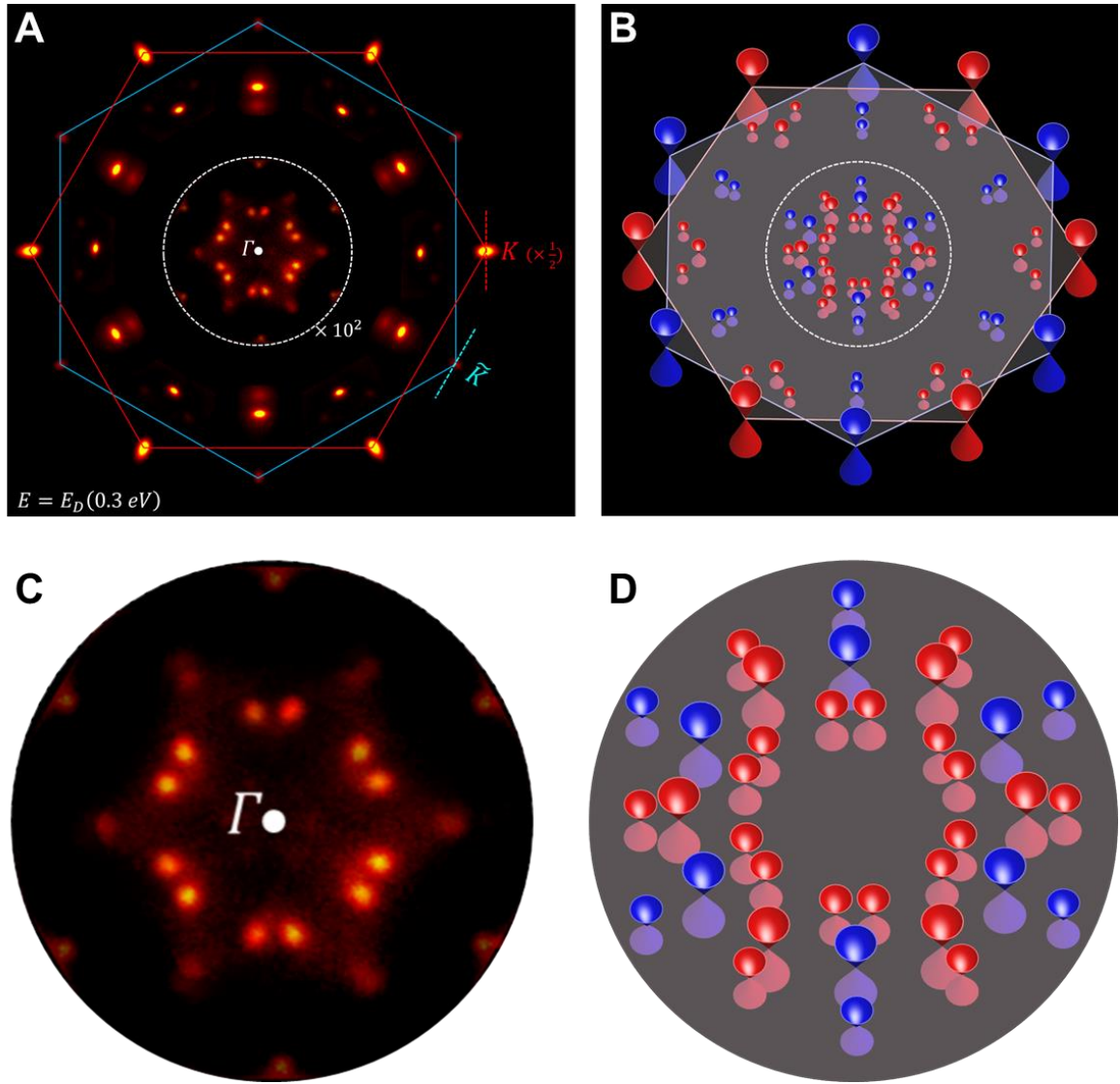

**FIGURE 2. Constant energy maps of ARPES spectra of graphene quasicrystal.** (A) A constant energy map at the Dirac point located at a binding energy of 0.3 eV, where the K points of the upper and lower layer graphene are $K$ and $\widetilde{K}$, respectively. The intensities near the $\Gamma$ point were magnified by $\times 10^2$ and the intensity of the Dirac cone at the $K$ point was reduced by half. (B) A schematic drawing of Dirac cones shown in (A), where red and blue Dirac cones come from the upper and lower layer graphene, respectively and the gray plane indicates the Dirac point so that the darker and brighter colors indicate binding energies above and below the Dirac point. (C) Enlarged constant energy maps near the $\Gamma$ point indicated by the white dotted circle in (A). (D) A schematic drawing of Dirac cones near the $\Gamma$ point shown in (C).

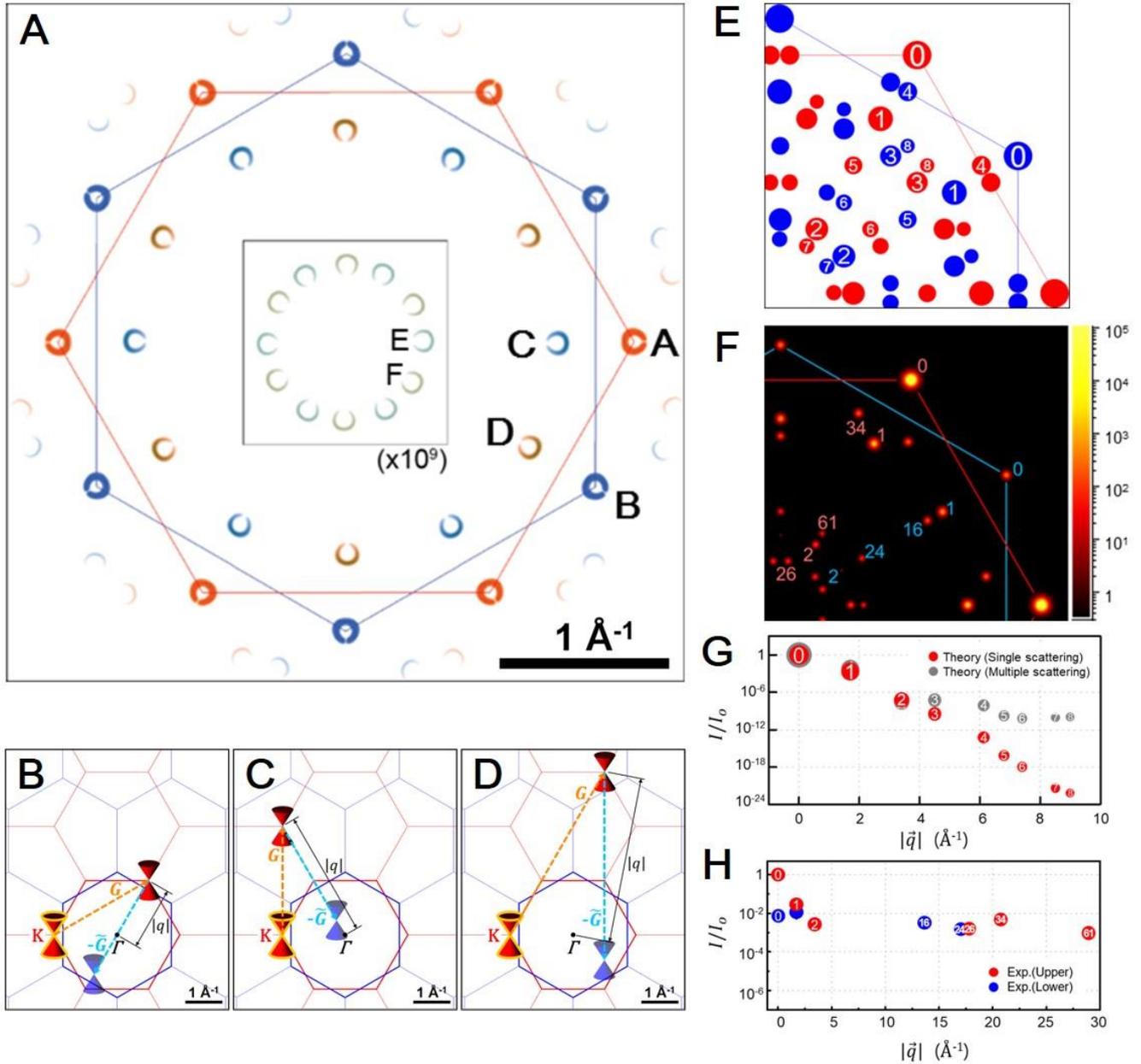

**FIGURE 3. Calculations of the electronic structure of graphene quasicrystal.** (A) A constant energy map at the Fermi energy (0.3 eV above the Dirac points) of simulated ARPES spectra of graphene quasicrystal. The color of each Dirac cone indicates whether the Dirac cone mainly originates from the states in the upper (red) or the lower (blue) graphene layer. The intensity of each Dirac cone indicates the scattering intensity plotted in a log scale. The intensities of the Dirac cones near the Γ point were magnified

by $\times 10^9$. (B-D) Umklapp scattering paths from the Dirac point of the upper layer with the shortest three wave vectors |**q**| involved in the Umklapp process (see text). Each panel shows the scattering involving different **G**. (E) A schematic drawing of the locations of Dirac cones in calculations, where the Dirac cones were ranked by the length of |**q**| and the size of circle is proportional to the intensity calculated theoretically (see Figs. S8 and S9 in SI also). (F) A schematic drawing of the locations of Dirac cones in ARPES measurements, where the Dirac cones were also ranked by the length of |**q**| and the intensity and size of circle is proportional to the intensity measured experimentally. (G) The theoretical calculations of the contribution to the ARPES intensity from the single (multiple) scattering plotted as a function of |**q**|. Red (grey) circles in (G) show the contribution from the single (multiple) scattering (see SI and Fig. S8 also). (H) The experimental intensities of Dirac cones plotted as a function of |**q**|.

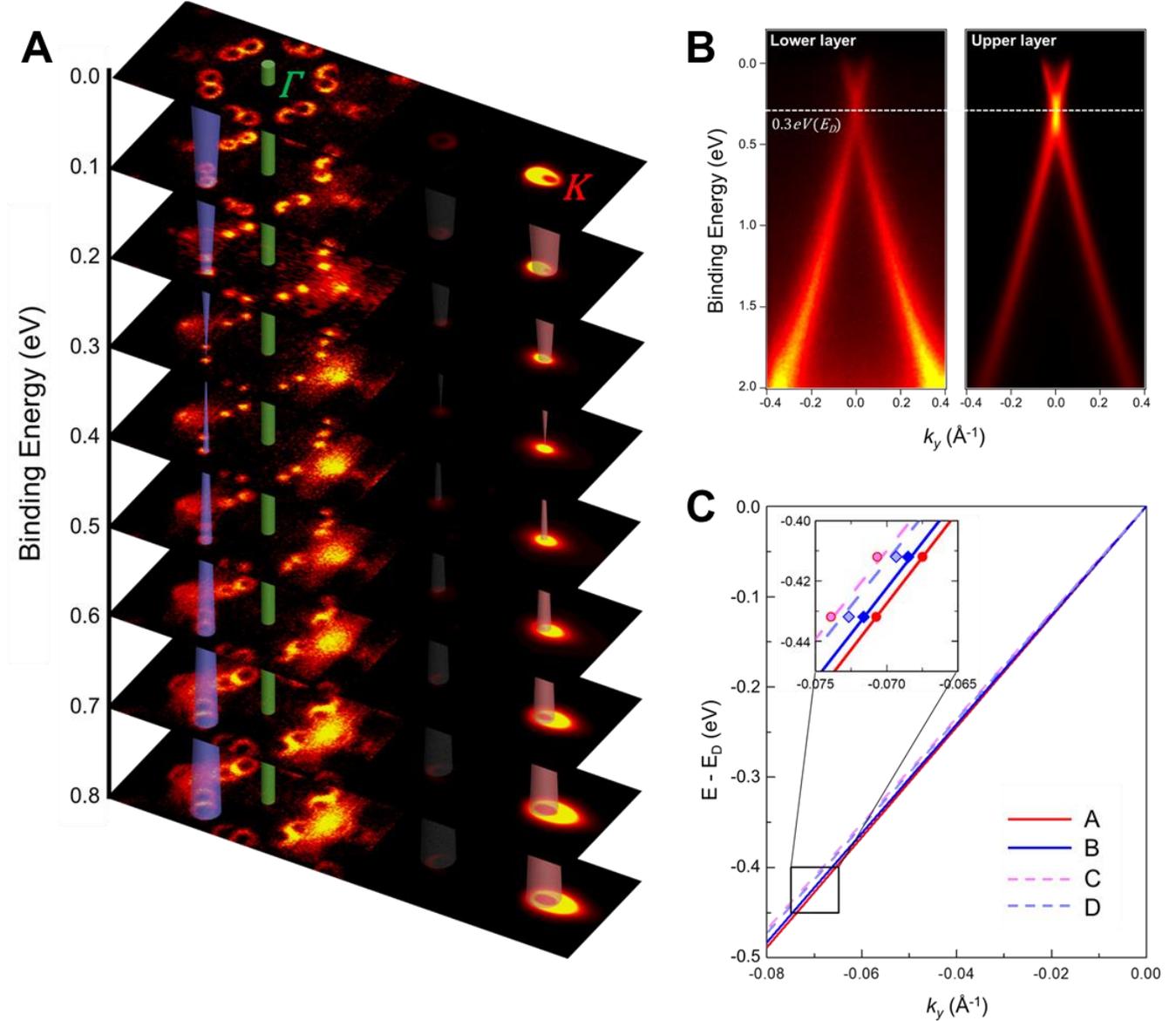

**FIGURE 4. ARPES spectra and Fermi velocities of graphene quasicrystal.** (A) Constant energy maps of ARPES spectra of graphene quasicrystal with different binding energies. (B) Energy-momentum dispersions of Dirac cones at the K points of the upper and lower layer graphene. (C) The fitted lines of energy-momentum dispersions of Dirac cones used to extract Fermi velocities (see Fig. S4F in SI also for detailed comparison between the experimental dispersions and the fitted lines), where the Fermi velocities of the A, B, C and D Dirac cones (see Fig. 3(A) for the notations of the Dirac cones) are $0.93 \times 10^6$,

$0.92\times10^6$, $0.89\times10^6$ and $0.91\times10^6$ m/sec, respectively. In the inset of (C), experimental dispersions are overlapped with the fitted lines.

# Supplementary Information

# Dirac Electrons in a Dodecagonal Graphene Quasicrystal

Sung Joon Ahn[1†], Pilkyung Moon[2†], Tae-Hoon Kim[3†], Hyun-Woo Kim[1], Ha-Chul Shin[1], Eun Hye Kim[1], Hyun Woo Cha[3], Se-Jong Kahng[4], Philip Kim[5], Mikito Koshino[6], Young-Woo Son[7]*, Cheol-Woong Yang[3]* and Joung Real Ahn[1,8]*

[1]*Department of Physics and SAINT, Sungkyunkwan University, Suwon, Republic of Korea*

[2]*New York University and NYU-ECNU Institute of Physics at NYU Shanghai, Shanghai, China*

[3]*School of Advanced Materials Science and Engineering, Sungkyunkwan University, Suwon, Republic of Korea*

[4]*Department of Physics, Korea University, Seoul, Republic of Korea*

[5]*Department of Physics, Harvard University, Cambridge, MA, USA*

[6]*Department of Physics, Osaka University, Machikaneyama, Toyonaka, Japan*

[7]*Korea Institute for Advanced Study, Seoul, Republic of Korea*

[8]*Samsung-SKKU Graphene Center, Sungkyunkwan University, Suwon, Republic of Korea*

[*]To whom correspondence should be addressed.

E-mail: jrahn@skku.edu(JRA), cwyang@skku.edu(CWY), hand@kias.re.kr (YWS)

[†]These authors contributed equally to this work.

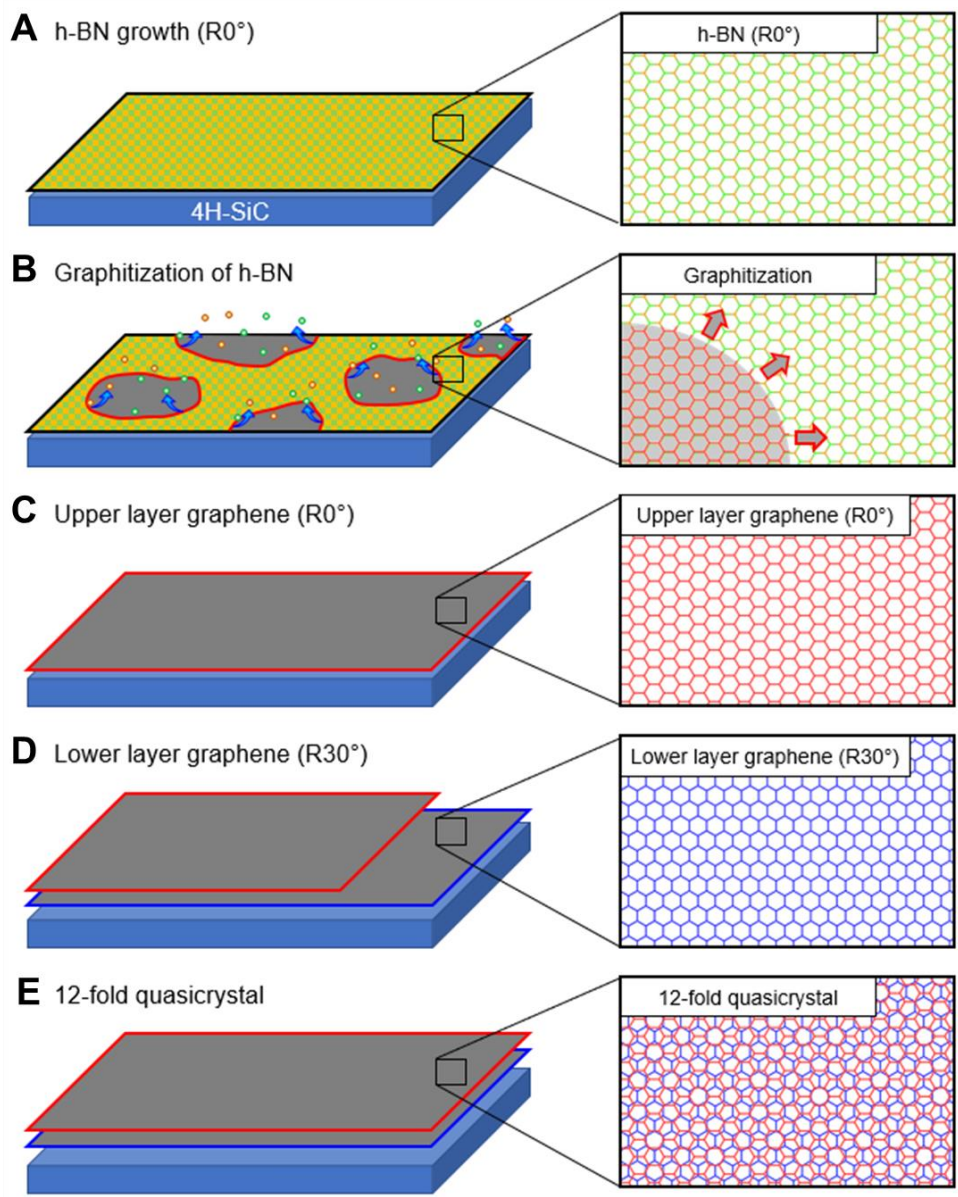

**FIGURE S1. Schematic drawings of graphene quasicrystal growth.** The graphene quasicrystal was obtained through two step epitaxial growths (see text). Corresponding atomic structure models of each process are shown in right side of each figure. (A) The growth of h-BN with R0º. (B) The change of h-BN with R0º into the upper layer graphene with R0º. (C) The complete growth of the upper layer graphene with R0º. (D) The growth of lower layer graphene with R30º between the upper layer graphene and the SiC substrate. (E) The complete growth of the graphene quasicrystal.

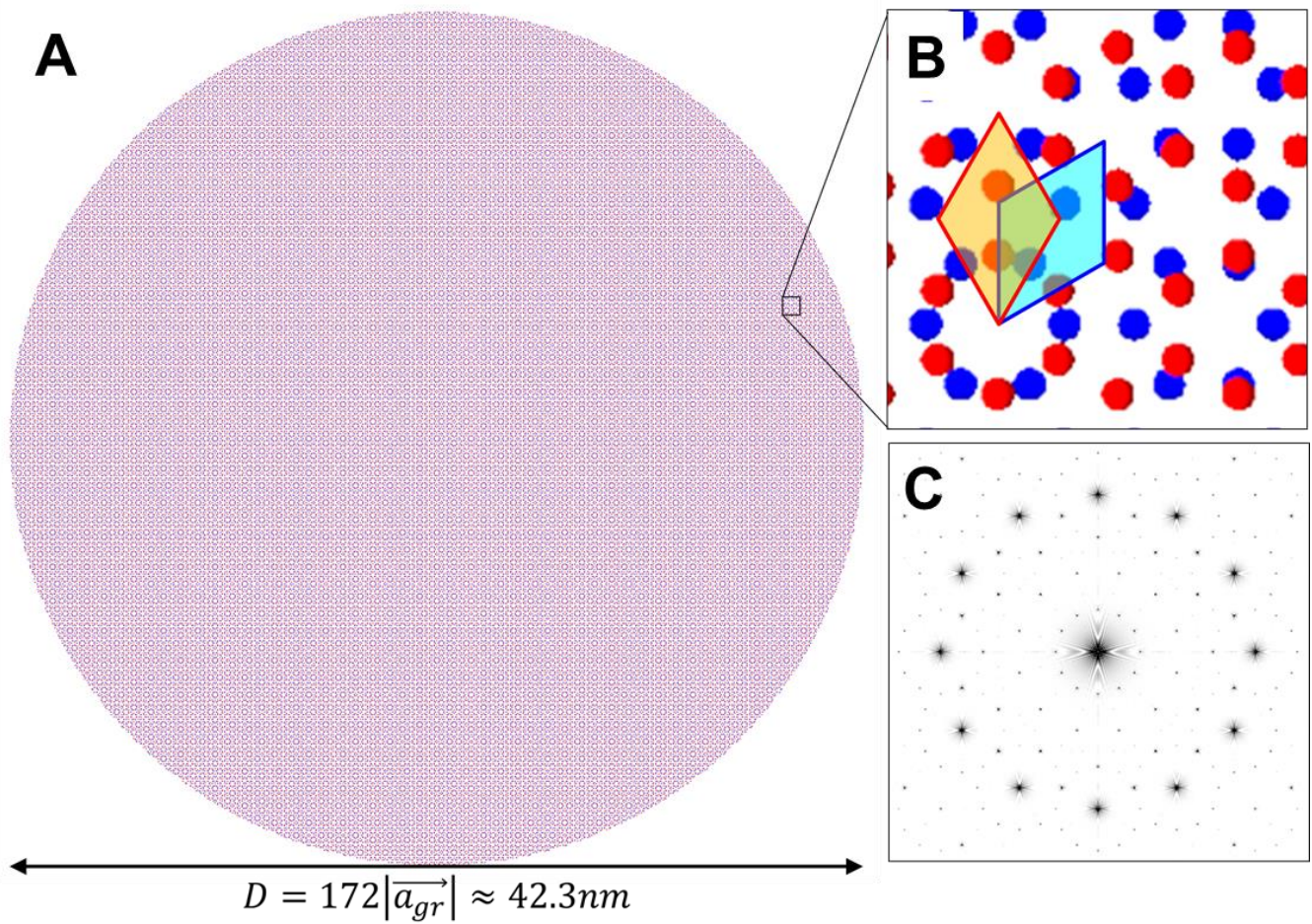

**FIGURE S2. Fourier transformation of graphene quasicrystal.** (A) An original large scale atomic structure model used to obtain the Fourier-transformed image in Fig. 1(B). (B) An enlarged atomic structure model of (A). Primitive unit cells of the upper and lower layer graphene are denoted by red and blue rhombuses, respectively. Diameter of the model in (A) is approximately 42.3 nm, which is 172 times the lattice constant of single layer graphene. (C) Fourier-transformed image obtained from (A).

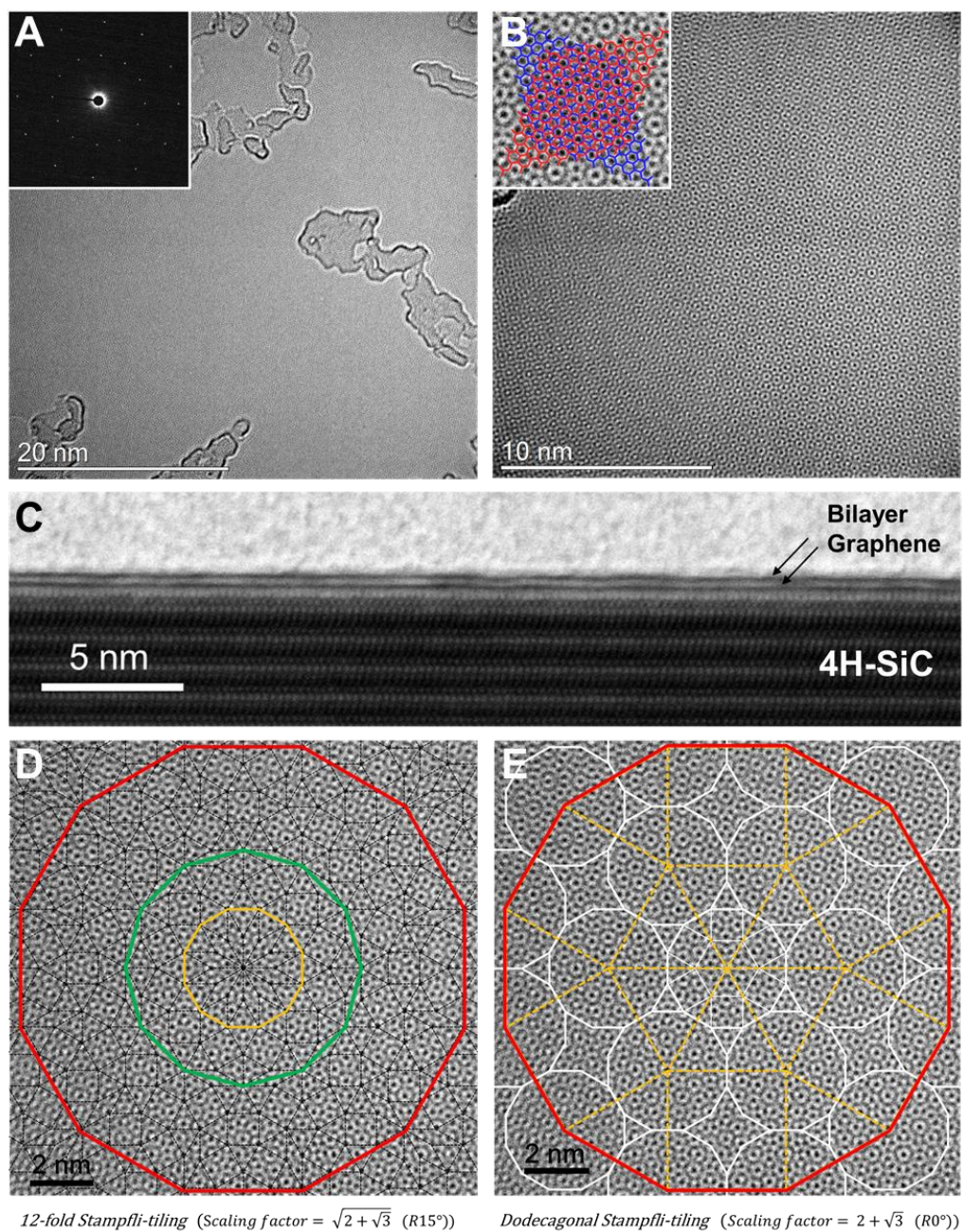

*12-fold Stampfli-tiling* (Scaling factor = $\sqrt{2+\sqrt{3}}$ (R15°))   *Dodecagonal Stampfli-tiling* (Scaling factor = $2+\sqrt{3}$ (R0°))

**FIGURE S3. TEM images of graphene quasicrystal.** The TEM images in (A)-(B) and (D)-(E) are measured after transferring graphene quasicrystal on a TEM grid. The cross-sectional TEM image in (C) is measured on SiC before transferring graphene quasicrystal. (A) A large-scale TEM image, where its selected area electron diffraction (SAED) pattern is shown in the inset. (B) An enlarged atomic scale TEM image, where the TEM image is overlapped with the atomic structure model of graphene quasicrystal in the inset and the red and blue atomic structure are the upper and lower layer graphene, respectively. (C) A cross-sectional TEM image of graphene quasicrystal on SiC, where bilayer graphene are clearly observed. (D)-(E) Two different Stampfli-inflation tiling, compared with the TEM image. (D) 12-fold inflation rule with a scaling factor of $\sqrt{2+\sqrt{3}}$, (E) Dodecagonal inflation rule with a scaling factor of $2+\sqrt{3}$.

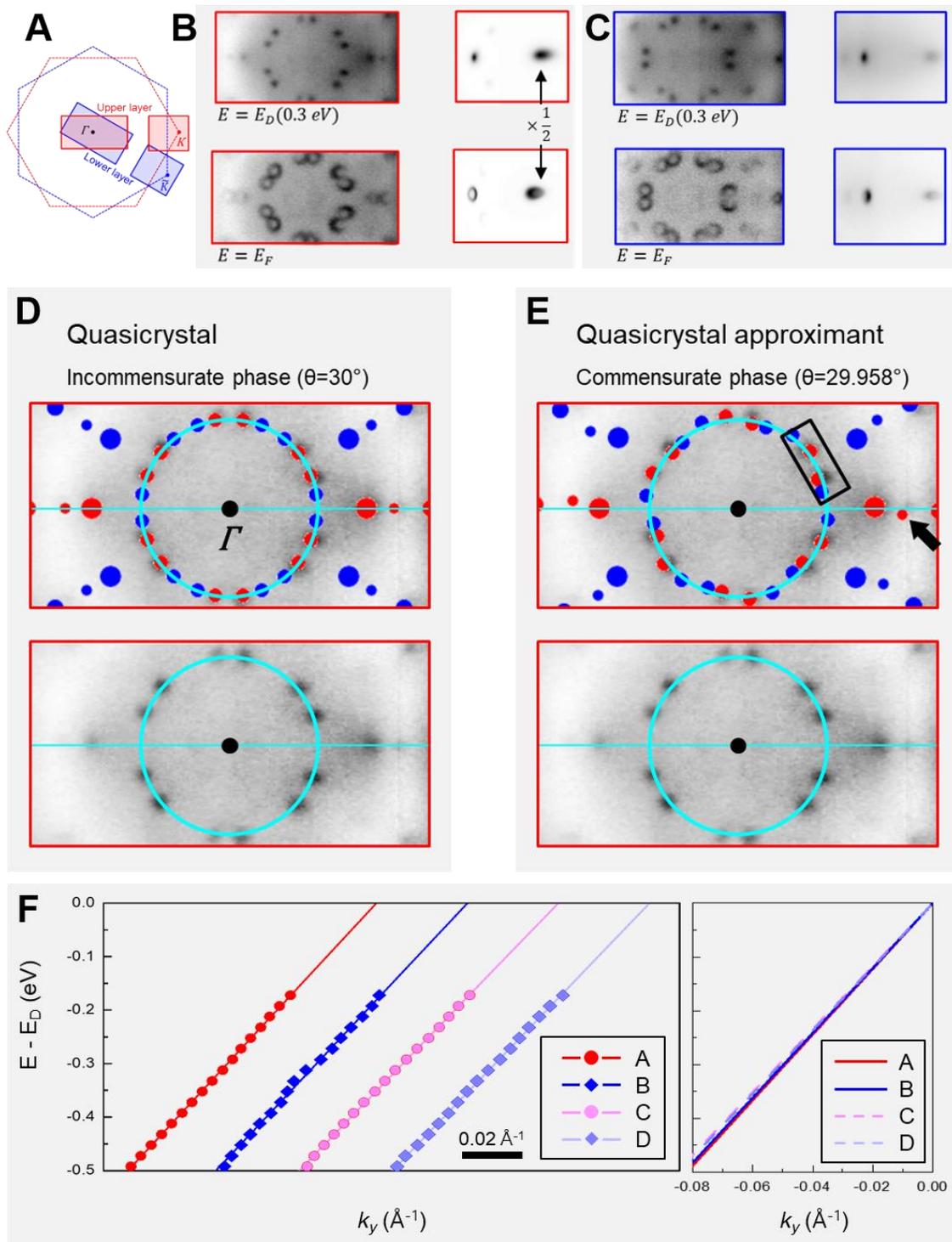

**FIGURE S4. Constant energy maps of ARPES spectra with different binding energies.** (A) A schematic drawing of momentum space measured. 1st Brillouin zones of upper and lower layer graphene

are denoted by red and blue hexagons, respectively. (B)-(C) Constant energy maps of (B) upper and (C) lower layer graphene measured with different binding energies, where $E_D$ and $E_F$ are the Dirac point and the Fermi energy, respectively. In (B), the intensity of the Dirac cone at the *K* point was reduced by half. (D)-(E) Comparison of (D) graphene quasicrystal with an exact rotational angle of 30° to (E) quasicrystal approximant with an approximant rotational angle of 29.958°. The Dirac cone replicas near the $\Gamma$ point of the quasicrystal approximant in (E) are not consistent with that measured from ARPES experiments, as indicated by the black rectangles and arrow. (F) In the left panel, the symbols are experimental energy-momentum dispersions of Dirac cones and the solid lines are fitted ones (see Fig. 3(A) for the notations of the Dirac cones). In (F), all dispersions were fit linearly and the fitted lines in Fig. 4(C) are displayed in the right panel.

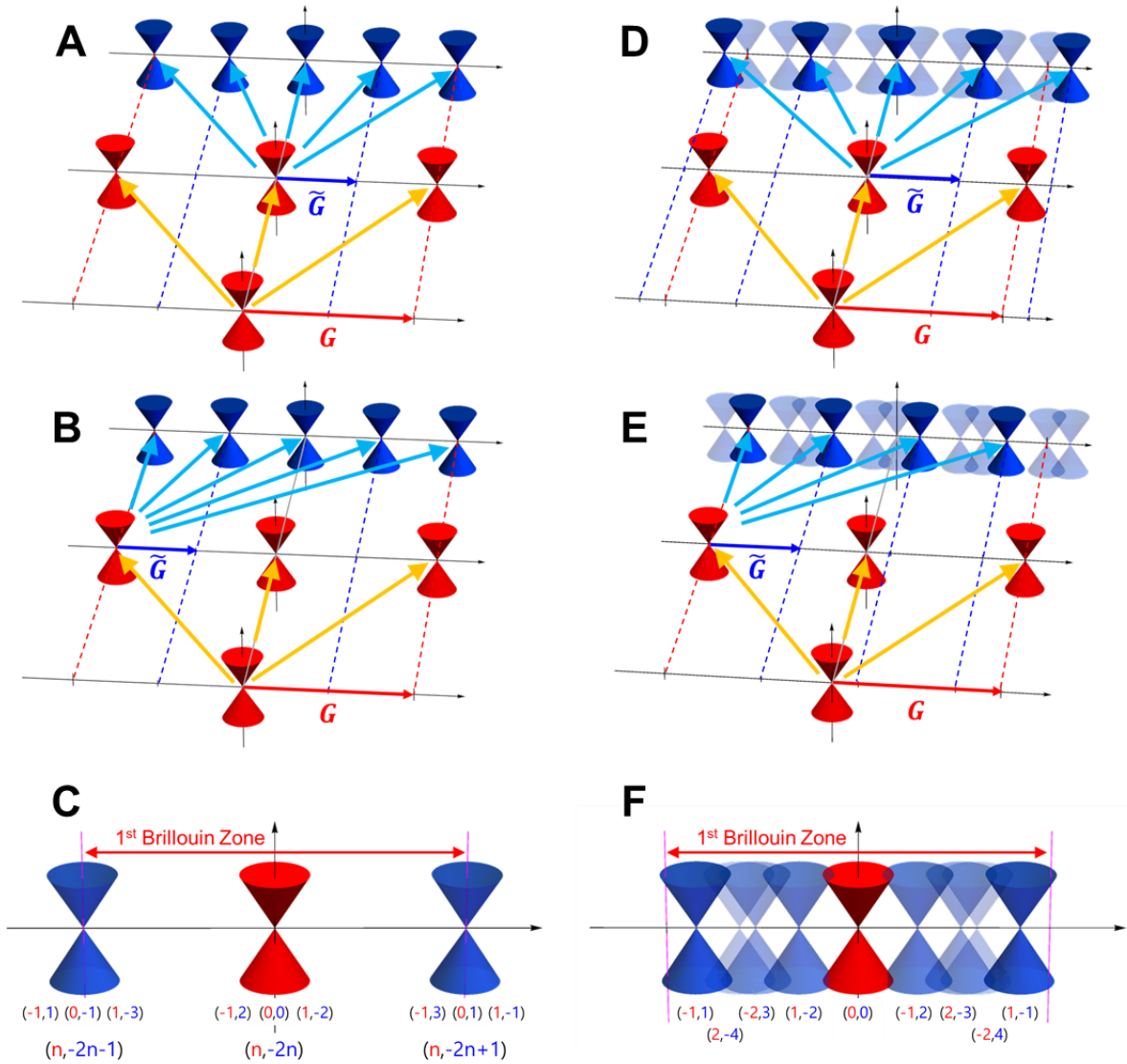

**FIGURE S5. Schematic drawings of Umklapp scattering of a Dirac cone in one-dimensional (1D) periodic and quasiperidoic order.** (A)-(C) Umklapp scattering in 1D periodic order, where two different reciprocal lattice vectors, $\mathbf{G}$ and $\widetilde{\mathbf{G}}$, are commensurate to each other. (A)-(B) show two different scattering paths. (C) Dirac cones formed by the the two different scatterting paths. (D)-(F) Umklapp scattering in 1D quasiperiodic order, where two different reciprocal lattice vectors, $\mathbf{G}$ and $\widetilde{\mathbf{G}}$, are incommensurate to each other. (D)-(E) show two different scattering paths. (F) Dirac cones formed by the the two different scatterting paths, where multiple Dirac cone replicas are created.

## Section 1. Theoretical calculations

### 1-1. Simulation of angle-resolved photoemission spectroscopy (ARPES) intensities

We use the Fermi golden rule to simulate ARPES spectra (*R1, R2*). The transition probability (*I*) from an initial Bloch state ($\Psi_i$) to an outgoing state ($\Psi_f$, approximated by a plane-wave state $e^{i\mathbf{p}\cdot\mathbf{r}}$) is calculated by

$$I(\mathbf{p},\omega) : \left|\langle\Psi_f|H_{ph-el}|\Psi_i\rangle\right|^2 \delta(\hbar\omega + E_i - E_f)$$

by using the dipole approximation for the photon-electron interaction Hamiltonian $H_{ph-el}$ ; $\mathbf{p}\cdot\mathbf{A}e^{i\mathbf{q}\cdot\mathbf{x}}$. Here $\mathbf{q}$ and $\omega$ are the momentum and frequency of incident photon, respectively, $E_i$ ($E_f$) is the energy of $\Psi_i$ ($\Psi_f$), and $\mathbf{p}$ is the momentum of outgoing electron.

### 1-2. Interlayer Hamiltonian for general incommensurate atomic layers

The interlayer interaction of the graphene quasicrystal couples a Bloch state $\mathbf{k}$ in the upper layer with the state $\tilde{\mathbf{k}}$ in the lower layer when a generalized Umklapp scattering process between incommensurately stacked atomic layers (*R3, R4*)

$$\mathbf{k}+\mathbf{G}=\tilde{\mathbf{k}}+\tilde{\mathbf{G}}$$

is satisfied. Here, $\mathbf{G} = m_1 \mathbf{a}_1^* + m_2 \mathbf{a}_2^*$ and $\tilde{\mathbf{G}} = \tilde{m}_1 \tilde{\mathbf{a}}_1^* + \tilde{m}_2 \tilde{\mathbf{a}}_2^*$ ($m_1, m_2, \tilde{m}_1, \tilde{m}_2 \in \mathbb{Z}$) run over all the reciprocal lattice vectors of the upper ($\mathbf{a}_1^*$, $\mathbf{a}_2^*$) and lower ($\tilde{\mathbf{a}}_1^*$, $\tilde{\mathbf{a}}_2^*$) layers, respectively. Since $\mathbf{G}$ and $\tilde{\mathbf{G}}$ are not commensurate, the scattering by the interlayer interaction

$$U = -\sum_{X,X'} T_{XX'}(\mathbf{R}_{X'} - \mathbf{R}_X) |\mathbf{R}_{X'}\rangle\langle\mathbf{R}_X| + \text{h.c.}$$

makes an infinite number of Dirac cone replicas in the Brillouin zone through the corresponding scattering strength

$$U_{XX'}(\mathbf{k}',\mathbf{k}) \equiv \langle \mathbf{k}', X'|U|\mathbf{k}, X\rangle = -\sum_{\mathbf{G},\mathbf{G}'} t_{XX'}(\mathbf{k}+\mathbf{G}) e^{-i\mathbf{G}\cdot\boldsymbol{\tau}_X + i\mathbf{G}'\cdot\boldsymbol{\tau}_{X'}} \delta_{\mathbf{k}+\mathbf{G},\mathbf{k}'+\mathbf{G}'}.$$

Here, $X = A, B$ ($X' = A', B'$) denotes the sublattice index of the upper (lower) layer,

$$\mathbf{R}_X = n_1 \mathbf{a}_1 + n_2 \mathbf{a}_2 + \boldsymbol{\tau}_X \quad (\text{layer 1}),$$
$$\mathbf{R}_{X'} = n'_1 \mathbf{a}'_1 + n'_2 \mathbf{a}'_2 + \boldsymbol{\tau}_{X'} \quad (\text{layer 2}),$$

for integers $n_i$ and $n'_i$ are the lattice positions of each layer, $\boldsymbol{\tau}_X$ and $\boldsymbol{\tau}_{X'}$ denote the sublattice positions inside the unit cell, and $t(\mathbf{q})$ is the wave vector $\mathbf{q}$ component of the in-plane Fourier transform of the interlayer transfer integral defined by

$$t_{XX'}(\mathbf{q}) = \frac{1}{\sqrt{SS'}} \int T_{XX'}(\mathbf{r} + z_{XX'}\mathbf{e}_z) e^{-i\mathbf{q}\cdot\mathbf{r}} d^2 r,$$

where $-T_{XX'}(\mathbf{R}_{X'} - \mathbf{R}_X)$ is the transfer integral from the site $\mathbf{R}_X$ to $\mathbf{R}_{X'}$. In TBG, we consider the single-orbital tight-binding model for $p_z$ orbital of carbon atom, and assume that the transfer integral between any two orbitals is written in terms of the Slater-Koster form as *(R5)*,

$$-T(\mathbf{R}) = V_{pp\pi}\left[1 - \left(\frac{\mathbf{R}\cdot\mathbf{e}_z}{R}\right)^2\right] + V_{pp\sigma}\left(\frac{\mathbf{R}\cdot\mathbf{e}_z}{R}\right)^2,$$
$$V_{pp\pi} = V_{pp\pi}^0 e^{-(R-a_0)/r_0},$$
$$V_{pp\sigma} = V_{pp\sigma}^0 e^{-(R-d_{G-G})/r_0}.$$

Here $\mathbf{e}_z$ is the unit vector perpendicular to the graphene plane, $a_0 \approx 0.142$ nm is the distance of neighboring A and B sites on graphene, and $d_{\text{G-G}} \approx 0.335$ nm is the interlayer spacing between graphene layers. The parameter $V^0_{pp\pi}$ is the transfer integral between the nearest-neighbor atoms on graphene, and $V^0_{pp\sigma}$ is the transfer integral between vertically located atoms on the neighboring layers. We take $V^0_{pp\pi} \approx -2.7\,\text{eV}$ and $V^0_{pp\sigma} \approx 0.48\,\text{eV}$ to fit the dispersions of monolayer graphene, and the decay length of the transfer integral $\delta$ as $0.319 a_0$ so that the next-nearest intralayer coupling becomes $0.1 V^0_{pp\pi}$ *(R6-R8)*. Since the transfer integral is isotropic along the in-plane direction, we can write $t(\mathbf{q}) = t(q)$ with $q = |\mathbf{q}|$. Dirac cone replicas resulting from larger scattering strength exhibit higher intensity in the calculated ARPES map.

Figure S6 shows one of the single Umklapp scattering process. In this example, the interlayer interaction couples the K point of the upper layer with the $\tilde{\mathbf{k}}$ point of the lower layer by $\mathbf{G} = \mathbf{a}_1^* + \mathbf{a}_2^*$ and $\tilde{\mathbf{G}} = \tilde{\mathbf{a}}_1^* + \tilde{\mathbf{a}}_2^*$. Since the amplitude of the coupling $t(q)$ for $q = |\mathbf{K} + \mathbf{G}|$ is 0.110 eV, and the energy difference $\Delta E$ between the states $K$ and $\tilde{\mathbf{k}}$ is about 3 eV, this scattering contributes to the ARPES intensity at $\tilde{\mathbf{k}}$ approximately by $|t(q)/\Delta E|^2 \sim 0.00136$ relative to the intensity of the main Dirac cone. The actual contribution to the intensity varies depending on the relative phase between the sublattices.

## 1-3. Wave functions involved in Umklapp process

In the incommensurately stacked crystals (e.g., TBG), the state $|\mathbf{k}\rangle$ in the upper layer interacts with infinitely many distinct states $|\tilde{\mathbf{k}}\rangle$ in the lower layer. The left, middle, and right panels of the first row in Fig. S7 show that the Umklapp scattering process from the Dirac point ($\mathbf{k} = \mathbf{K}$) of the upper layer by three different reciprocal lattice vectors $\mathbf{G} = \mathbf{a}_1^* + \mathbf{a}_2^*$, $\mathbf{a}_2^*$, $\mathbf{a}_1^* + 2\mathbf{a}_2^*$, respectively. We can see that the scattered wave vectors $\mathbf{q} = \mathbf{k} + \mathbf{G}$ are mapped to distinct $\tilde{\mathbf{k}}$ ($\equiv \mathbf{k} + \mathbf{G} \pmod{\tilde{\mathbf{G}}}$) of the lower layer, since the reciprocal vectors of the upper and lower layers ($\mathbf{G}$ and $\tilde{\mathbf{G}}$) are neither identical nor commensurate.

We can also check this from the wave functions of the corresponding wave vectors. Each panel in the second row shows the lattice structure of the upper layer (showing the A-sublattice site only) together with the phase of the corresponding wave function $|\mathbf{k}+\mathbf{G}\rangle$ at the lattice points. The intensity map in the background shows the phase of the plane wave with a corresponding wave vector $\mathbf{k} + \mathbf{G}$. Although the spatial phase pattern of the plane wave varies with $\mathbf{G}$, the phase at the lattice points are invariant over different $\mathbf{G}$ due to the periodicity of the lattice. Thus, the wave functions $|\mathbf{k}+\mathbf{G}\rangle$ of the upper layer with different reciprocal vector $\mathbf{G}$ are equivalent (Bloch theorem).

By the generalized Umklapp process, $|\mathbf{k}+\mathbf{G}\rangle$ of the upper layer interacts with $|\tilde{\mathbf{k}}+\tilde{\mathbf{G}}\rangle$ of the lower layer, where $\mathbf{k}+\mathbf{G} = \tilde{\mathbf{k}}+\tilde{\mathbf{G}}$. Each panel in the third row in Fig. S7 shows the lattice structure of the lower layer (showing the A-sublattice site only) together with the phase of the corresponding wave function $|\tilde{\mathbf{k}}+\tilde{\mathbf{G}}\rangle$ at the lattice points. Note that, for each column, the intensity map in the second and third rows are identical, because of the momentum conversation $\mathbf{k}+\mathbf{G} = \tilde{\mathbf{k}}+\tilde{\mathbf{G}}$ during the Umklapp process. Again, we can see that the phase of the plane wave with the corresponding wave vector $\tilde{\mathbf{k}}+\tilde{\mathbf{G}}$ is consistent with the phase of the wave function at the lattice points. However, unlike the upper layer, the wave functions of the lower layer are not equivalent over different $\mathbf{G}$. Thus, the states $|\mathbf{k}+\mathbf{G}\rangle$ of the upper layer with different $\mathbf{G}$ interact with the states of the lower layer with distinct $\tilde{\mathbf{k}}$ (in modulus of $\tilde{\mathbf{G}}$).

**1-4. Dirac cone replicas from single scattering and their intensities**

Figure S8(a) shows the Brillouin zones of the two layer in the extended zone scheme, where the symbols represent the single scattering $\mathbf{k} + \mathbf{G}$ for some particular $\mathbf{k}$ (here chosen as the zone corner K) of the upper layer with several $\mathbf{G}$'s. Figure S8(b) plots the corresponding positions $\tilde{\mathbf{k}} = \mathbf{k} + \mathbf{G} - \tilde{\mathbf{G}}$ of these scattering inside the first Brillouin zone of the lower layer, which are coupled to $\mathbf{k}$ of the upper layer by the interlayer interaction. Figures S8(c) and the red circles in Fig. S8(d) show the amplitude of the coupling $t(q)$ and the contribution to the ARPES intensity $|\,t(q)\,/\,\Delta E\,|^2$ of each symbol, respectively. The scattering intensity becomes exponentially weaker for larger $q$. We plot the scattering pattern from every Dirac points of the two layers in Fig. S9. The strongest scattering contributes to the Dirac cone replicas near the Brillouin zone boundary (circles with edge lines) by an order of $10^{-3}$ relative to the intensity of the main Dirac cone, while the second strongest scattering contributes to the replicas near the Brillouin zone center by an order of $10^{-6}$.

**1-5. Dirac cone replicas from multiple scattering**

In addition to the single Umklapp scattering, there are infinitely many multiple scatterings, which map to the same spots in Fig. S9 but contribute additional intensities. In general, $n$ successive scatterings, which satisfy

$$\mathbf{k}^{(i)} + (-1)^{i+1}\mathbf{G}^{(i)} = \tilde{\mathbf{k}}^{(i)} + (-1)^{i+1}\tilde{\mathbf{G}}^{(i)}$$

with $\mathbf{G}^{(i)} = m_1^{(i)}\mathbf{a}_1^* + m_2^{(i)}\mathbf{a}_2^*$, $\tilde{\mathbf{G}}^{(i)} = \tilde{m}_1^{(i)}\tilde{\mathbf{a}}_1^* + \tilde{m}_2^{(i)}\tilde{\mathbf{a}}_2^*$ for each $i$-th scattering step, couple $\mathbf{k}^{(1)}$ of the upper layer with $\tilde{\mathbf{k}}^{(n)} = \mathbf{k}^{(1)} + m_1 \mathbf{G}_1^M + m_2 \mathbf{G}_2^M$ of the lower layer for an odd $n$ and $\mathbf{k}^{(n)} = \mathbf{k}^{(1)} + \tilde{m}_1 \mathbf{G}_1^M + \tilde{m}_2 \mathbf{G}_2^M$ of the same (upper) layer for an even $n$. Here,

$$m_1 = \sum_{i=1}^{n} m_1^{(i)}, \quad m_2 = \sum_{i=1}^{n} m_2^{(i)}, \quad \tilde{m}_1 = \sum_{i=1}^{n} \tilde{m}_1^{(i)}, \quad \tilde{m}_2 = \sum_{i=1}^{n} \tilde{m}_2^{(i)},$$

$\mathbf{G}_1^M \equiv \mathbf{a}_1^* - \tilde{\mathbf{a}}_1^*$, $\mathbf{G}_2^M \equiv \mathbf{a}_2^* - \tilde{\mathbf{a}}_2^*$, and

$$\begin{cases} \tilde{\mathbf{k}}^{(i)} = \tilde{\mathbf{k}}^{(i+1)} & (\text{odd } i) \\ \mathbf{k}^{(i)} = \mathbf{k}^{(i+1)} & (\text{even } i) \end{cases}.$$

For example, Figure S10 shows two examples which couples the two states in Fig. S6 (the K state of the upper layer and the $\tilde{\mathbf{k}} = K + \mathbf{G}_1^M + \mathbf{G}_2^M$ state of the lower layer) by triple scattering. Figures S10(a) and S10(b) show the examples for $\{(m_1^{(1)}, m_2^{(1)}), (m_1^{(2)}, m_2^{(2)}), (m_1^{(3)}, m_2^{(3)})\}$ of $\{(2,1), (-1,-1), (0,1)\}$ and $\{(0,0), (1,1), (0,0)\}$, respectively. The contribution to the ARPES intensity at $\tilde{\mathbf{k}}$ from the multiple scattering in Figs. S10(a) and S10(b) is obtained by the multiple of the scattering amplitude of each step involved in the scattering and it gives an order of $10^{-6}$ and $10^{-9}$, respectively. The ARPES intensity in Fig. 3(a) and the scattering amplitude with grey color in Figs. 3(g) and S8(d) show the contribution from every possible scatterings, including single and multiple scatterings. It is noteworthy that the contribution from the multiple scatterings become dominant for the Dirac cone replicas with large |**q**|.

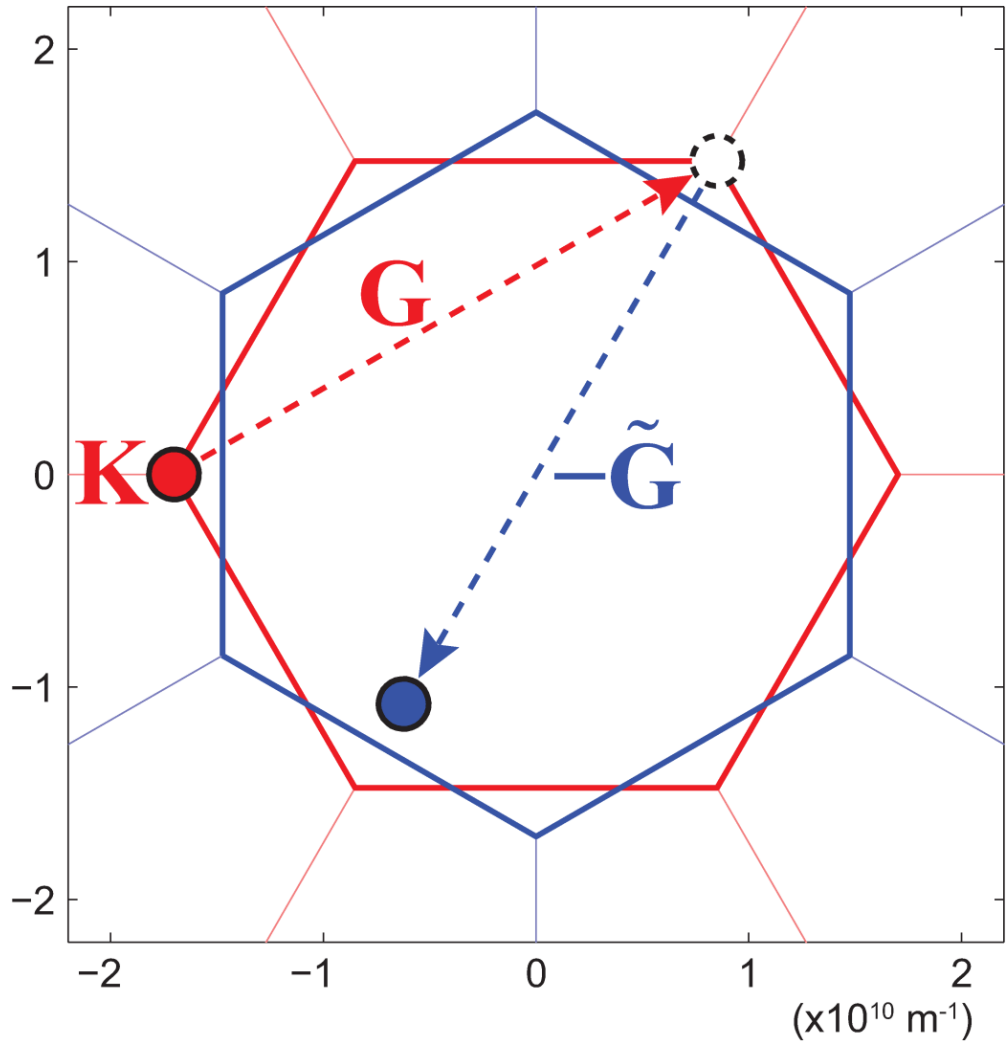

**FIGURE S6. Example of the single Umklapp scattering process.** The hexagon with a thick red (blue) line shows the first Brillouin zone of the upper (lower) graphene layer. The interlayer interaction couples the state in the upper layer (red circle) with the state in the lower layer (blue circle) with an assist of the reciprocal vectors of each layer, $\mathbf{G}$ and $\tilde{\mathbf{G}}$. This example shows the scattering with $\mathbf{G} = \mathbf{a}_1^* + \mathbf{a}_2^*$ and $\tilde{\mathbf{G}} = \tilde{\mathbf{a}}_1^* + \tilde{\mathbf{a}}_2^*$.

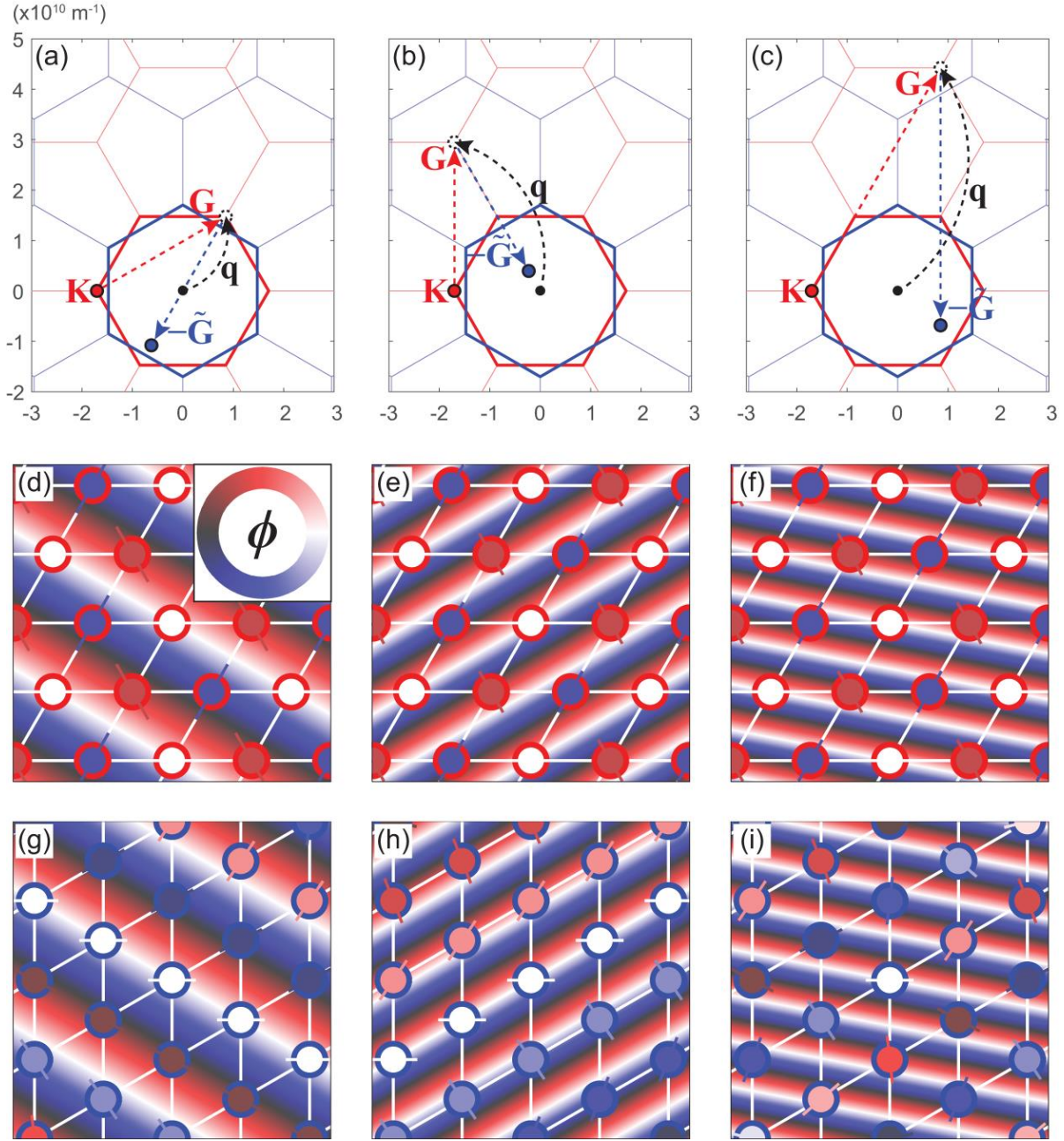

**FIGURE S7. Umklapp scattering and wave functions.** (a), (b), (c) Umklapp scattering paths involving three different **G** from the Dirac point of the upper layer. (d), (e), (f) Phase of wave function at each lattice point of the upper layer (showing A-sublattice sites only) together with the phase of plane wave (intensity map) for the wave vector **k**+**G** in (a), (b), (c), respectively. (g), (h), (i) Plots similar to (d), (e), (f) for the lower layer.

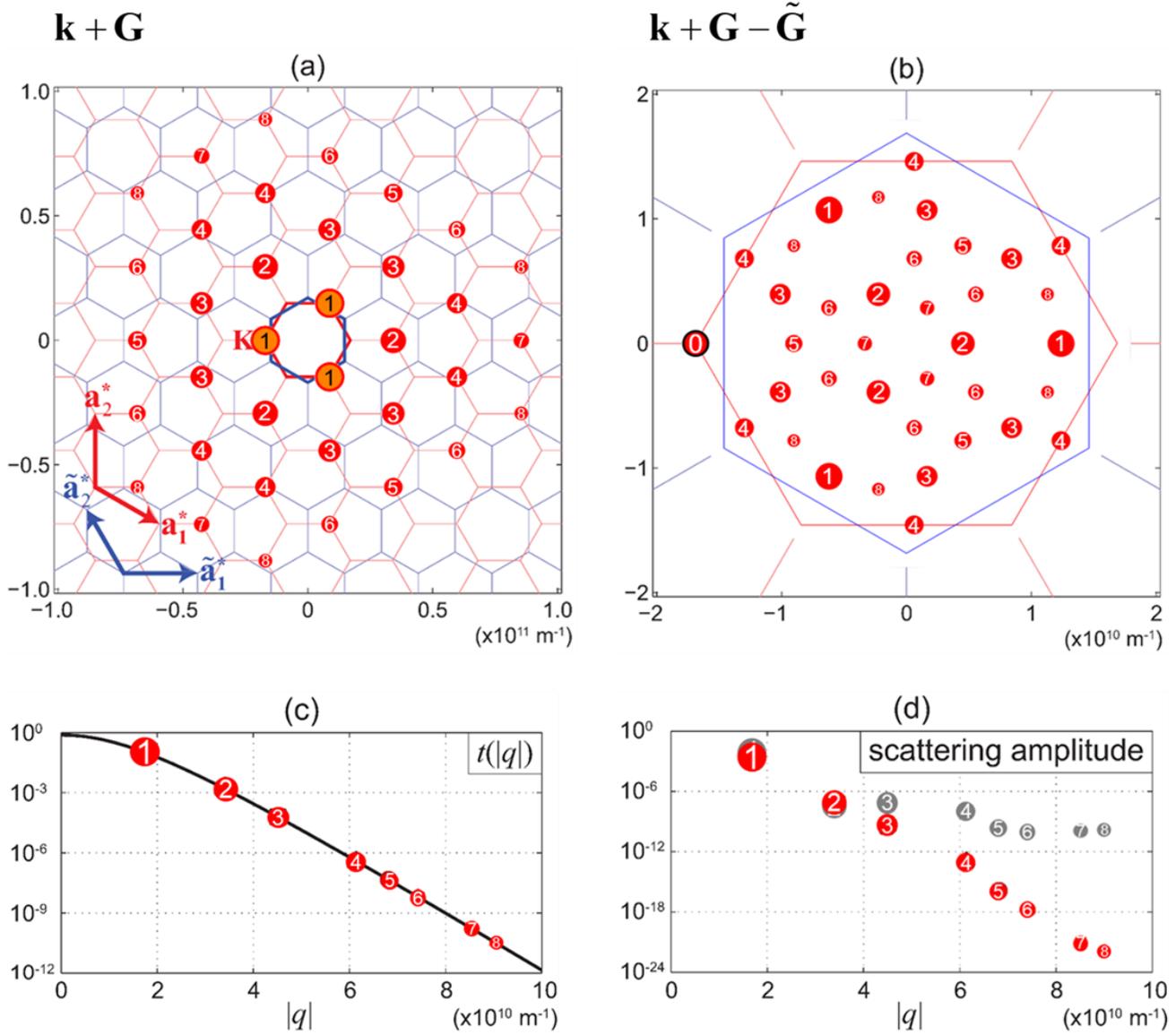

**FIGURE S8. Single scattering from a single K point of the upper layer.** The $\tilde{\mathbf{k}}$ states of the lower layer coupled to the K state in (a) the extended zone scheme and (b) the first Brillouin zone of the lower layer. (c) The amplitude of each single scattering, and (d) the contribution to the ARPES intensity. Red (grey) circles in (d) show the contribution from the single (multiple) scattering.

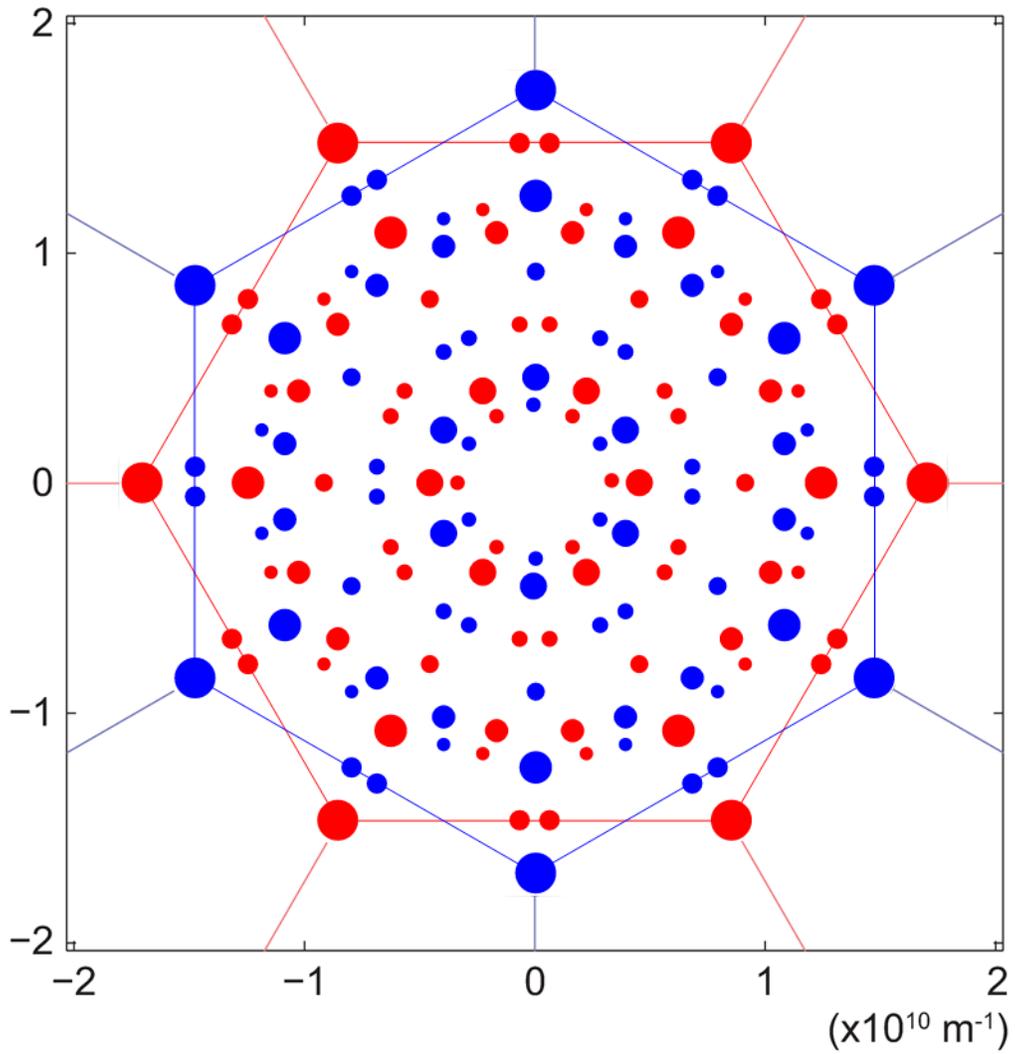

**FIGURE S9. Scattering pattern from every Dirac points of the two layers.** Plot similar to Fig. S8(b) for the scattering from every Dirac points of the two layers.

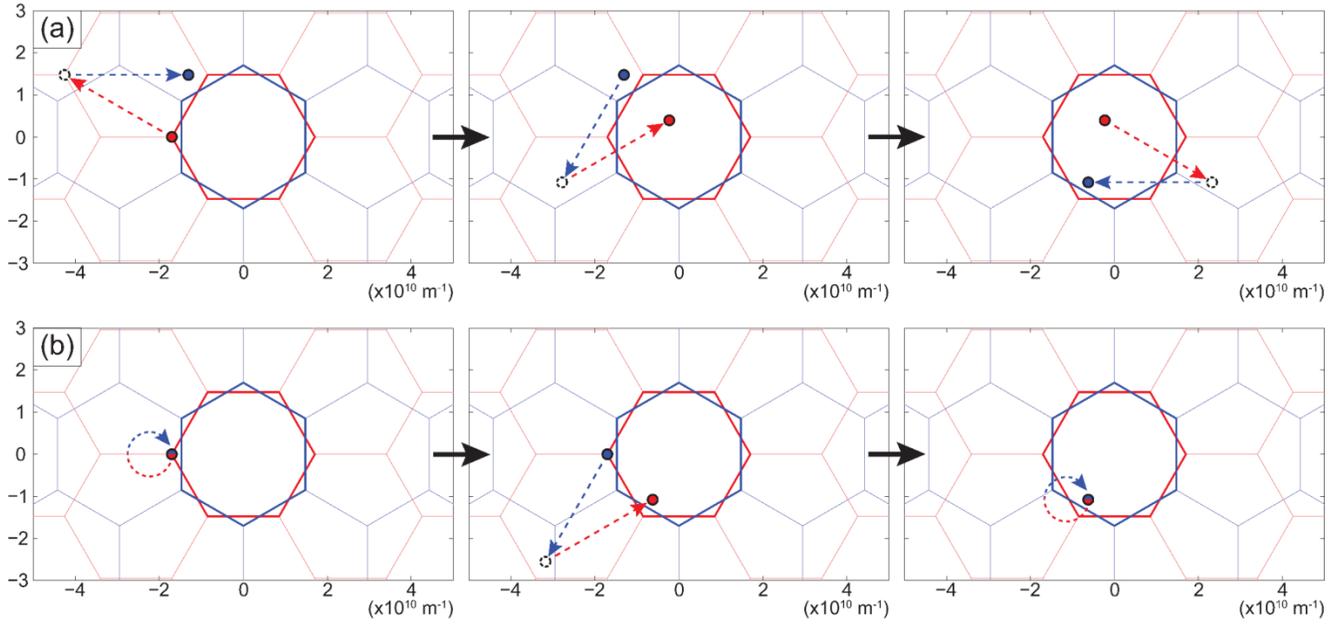

**FIGURE S10. Examples of multiple scattering between the states shown in Fig. S6.** The scattering for the scattering path $\{(m_1^{(1)}, m_2^{(1)}), (m_1^{(2)}, m_2^{(2)}), (m_1^{(3)}, m_2^{(3)})\}$ of (a) $\{(2,1), (-1,-1), (0,1)\}$ and (b) $\{(0,0), (1,1), (0,0)\}$. See text.

**Section 2. Experimental methods**

**2-1. ARPES experiments**

High-resolution ARPES constant energy maps and energy-momentum dispersions have been acquired with a commercial angle-resolved photoelectron spectrometer (R3000, VG Scienta) and monochromated He-Iα radiation (hν = 21.2 eV, VUV5k, VG Scienta) at room temperature. The resolutions of energy and angle (momentum) were 10 meV and 0.05° (~0.002 Å$^{-1}$), respectively.

**2-2. TEM experiments**

The plan-view TEM sample was prepared by using a Ni adhesive-stressor layer and the thermal release tape (*R9*). The first 30 nm Ni protection thin film is deposited by electron beam evaporation with 3 Å/s of evaporating speed to protect graphene layer from being damaged by a following sputtering process. The second Ni layer is deposited using a sputtering system with 200 nm of thickness. The twisted bilayer graphene was exfoliated and transferred onto the Si wafer which has thick $SiO_2$ layer on surface by using a thermal released tape (Haeun Chemtec, RP70N5). The $SiO_2$ layer was etched by hydrofluoric acid and floated Ni/graphene was picked up by a Quantifoil TEM grid (TedPella, #657-200-Au) and the Ni layer was etched by $FeCl_3$ based solution. Finally, the TEM sample was rinsed with D.I. water and the twisted bilayer graphene was successfully transferred on a TEM grid.

The cross-sectional TEM sample was fabricated using focused ion beam (JIB-4601F, JEOL) and the lift-out approach. A protective layer was gently deposited on the surface and low-kV Ga ion milling have been performed to prevent the damage on the twisted bilayer graphene during ion milling.

TEM images were taken using JEM-ARM200F (JEOL) operated at 80 kV of accelerating voltage. High resolution TEM imaging was performed with a low beam current of ~ 10 pA to prevent the likelihood of knock-on damage. The images were acquired during around 2.56 s with drift correction using OneView camera (Gatan).

Scanning transmission electron microscope (STEM) imaging was conducted with an aberration corrected STEM (JEM-ARM200F, JEOL) operated at 80 kV. STEM images were collected using a convergence semi-angle of 21 mrad and bright-field detector with a collection semi-angle of 0-17 mrad.

**SUPPLEMENTARY REFERENCES**